# Simple methods for converting equations between the SI, Heaviside-Lorentz and Gaussian systems

Paul Quincey, Independent researcher, Teddington, Middlesex, United Kingdom
paulgquincey@gmail.com

**Abstract**

School and undergraduate students are almost always taught the equations of electromagnetism using a set of conventions that are described as the SI. More advanced students are often introduced to different conventions that produce different equations for the same relationships, using either the Gaussian or Heaviside-Lorentz systems. In general, the connection between these equations is not simple. However, if the basis of each system is understood, conversion from SI equations to either Gaussian or Heaviside-Lorentz ones is very straightforward. The reverse processes are less straightforward, but more comprehensible when the fundamental differences are understood. Simple methods for these processes are presented, using a novel application of dimensional analysis, without factors of $\varepsilon_0^{1/2}$ or $(4\pi\varepsilon_0)^{1/2}$ appearing. It is also shown that when different physical quantities are given different symbols, and these are used consistently, the SI can be seen to provide general equations, with the Gaussian and Heaviside-Lorentz ones being simplifications of them. This removes any need for 'system-independent' versions of electromagnetic equations, with additional parameters that take different values in the different systems, which have been proposed in various forms over many decades.

## 1      Background

1.1 Historical background to the SI, Heaviside-Lorentz and Gaussian systems

The history of unit systems for electromagnetism is long and fascinating [1-3], but will not be described here in detail. Some key dates were:

- 1832, when Carl Friedrich Gauss found an ingenious way to measure the intensity of the Earth's magnetic field using only units for mass, length, and time;

- 1846, when Gauss's colleague Wilhelm Weber showed that alongside the unit system used by Gauss, which was based on magnetic phenomena, there was an analogous, different system based on electrical ones. The two systems were later called 'electromagnetic' and 'electrostatic';

- 1873, when a committee including James Clerk Maxwell[1] and William Thomson (later Lord Kelvin) endorsed Gauss and Weber's "three base unit" idea and proposed standardising the units as the centimetre, gram and second (CGS). They did not choose between the two equally valid systems, CGS-electrostatic and CGS-electromagnetic, which had different units for each electrical and magnetic quantity (the statcoulomb and abcoulomb for charge, for example), and which used equations whose terms differed by various factors of *c*;

- 1882, when Hermann von Helmholtz [4] advocated an approach that became a hybrid of the two CGS systems, with a single set of units and a single set of equations, which he called "Gauss's system".

---

[1] Apart from the long shadow cast by Gauss, and an instinct to 'keep it simple', perhaps Maxwell was influenced by his pioneering work on colour photography, involving three primary colours.

His main aim was to give magnetic pole strength the same dimensions as electric charge, as suggested by Gauss. This system, which is described in more detail below, was developed further in the 1880s by Helmholtz's former student Heinrich Hertz;

1901, when Giovanni Giorgi proposed introducing an electrical fourth base unit, which naturally leads to a single set of units and equations, and also changing the other base units to the metre, kilogram and second;

1902, when Hendrik Lorentz published an encyclopaedia article on electromagnetism [5] using a "rationalised" version (a term explained below) of the Gaussian system. As Oliver Heaviside had been a prominent advocate for rationalisation in the 1890s, this became known as the Heaviside-Lorentz system. Lorentz referred to the system we know as Gaussian as the "mixed electrostatic-electromagnetic" system, which he attributed to Gauss, Helmholtz and Hertz;

1960, when Giorgi's proposals were formally included as a key part of the newly created International System of Units or SI.

One obvious point is that the process has continued over many generations of scientists, making the narrative hard to follow. Another is that, more than a century ago, there was a fundamental divergence between three-base-unit systems such as Gaussian and Heaviside-Lorentz, and the four-base-unit approach that led to the SI. These paths have continued separately ever since.

1.2    Some previous treatments of the inter-relationship between these systems

Because the three systems use different equations to describe the same physical relationships, there has been a long tradition of creating overarching equations with extra parameters that take different values in the various systems. This can be done in many ways, some of which are indicated below. As set out in Section 2 below, once it is realised that symbols such as $\mathbf{B}$, $\mathbf{H}$, and $\mathbf{M}$ refer to different quantities in the different systems, the only difference is that the dimensional, experimentally-determined constant $\varepsilon_0$ in the SI is given a fixed numerical value in the other systems. In effect, the SI is already the general, overarching system, and no extra parameters are needed.

The Appendix on Units and Dimensions in John David Jackson's *Classical Electrodynamics* [6] is admirably clear and rightly well-known, and is based on the extra parameter approach. His generalized equations contain four extra parameters: $k_1$, $k_2$, $k_3$ and $\alpha$, where $k_1 = c^2 k_2$ and $k_3 = 1/\alpha$. The main flaw is that the entries for $\varepsilon_0$ and $\mu_0$ in the Appendix's Table 2 are defined as $\mathbf{D}/\mathbf{E}$ and $\mathbf{B}/\mathbf{H}$ respectively (when $\mathbf{P}$ and $\mathbf{M}$ are zero), without regard to the different definitions of $\mathbf{B}$, $\mathbf{H}$ and $\mathbf{D}$ in the different systems, which are described in Section 2 below. The symbols $\varepsilon_0$ and $\mu_0$ in that Table therefore do not represent what they do within SI equations, their only natural habitat. For example, the Table states that both $\varepsilon_0$ and $\mu_0$ are equal to 1 within both the Gaussian and Heaviside-Lorentz systems. If $\varepsilon_0$ and $\mu_0$ have their standard definitions, this much-repeated slip would imply that $c = 1$, and therefore that $c$ could be removed from all Gaussian and Heaviside-Lorentz equations.

Other examples of the extra parameter approach are Gelman [7] and a more recent paper in this journal by Dietmar Petrascheck [8]. Gelman's generalized equations contain five extra parameters: $f_e$, $K$, $f_m$, $\alpha_1$ and $\alpha_2$.

The hierarchical nature of unit systems, from "general" to "simplified" and ultimately to systems of "natural units", is described in a paper by Quincey and Brown [9]. The "extra parameter" approach takes the view that the differences between unit systems are arbitrary choices, without a hierarchy, a view which creates needless complexity and hinders the task of converting equations from one system to another.

In terms of providing instructions for converting equations from one system to another, Appendix D entitled 'SI and Gaussian formulas' in Edward Purcell and David Morin's *Electricity and Magnetism* [10] does an excellent job of explaining how to convert from SI to the Gaussian system, without factors of $(4\pi\varepsilon_0)^{1/2}$, using an approach very similar to the one presented below in Table 2. Equivalent tables for the same conversion can be seen in the textbooks by Philip Clemmow [11] and William Duffin [12], specifically Tables A.3 and C.2 respectively. The current paper extends this approach to the Heaviside-Lorentz system, and to the more difficult task of converting equations in the other direction.

Appendix B, entitled 'Gaussian Units', in Andrew Zangwill's *Modern Electrodynamics* [13], and a more recent paper in this journal by Anupam Garg [14] present methods for converting equations between the SI and Gaussian systems in both directions. The methods set out here in Tables 2 and 7 are simpler than the ones in these references, where factors of $(4\pi\varepsilon_0)^{1/2}$ are used for the conversion in both directions.

Section 2 describes the three systems, starting with the SI, followed by Heaviside-Lorentz, and then Gaussian, even though this was not the order historically, for reasons that should become clear.

## 2    Descriptions of the three systems

### 2.1    The SI

By having four base units, the SI starts from the position that electric charge cannot be seen as a combination of mass, length and time. Whether or not we agree with this position – and it seems very sensible to this author[2] – we can easily move from four to three base units if we want to. In effect, this was done in 1873 by setting either $4\pi\varepsilon_0$ equal to 1 (to make CGS-electrostatic) or $\mu_0/4\pi$ equal to 1 (to make CGS-electromagnetic), where $\varepsilon_0$ and $\mu_0$ are defined in the next paragraph. Both constants are thus removed from equations, while a base unit is removed from the unit system [9]. This is analogous to setting $c$ equal to 1 in systems of "natural" units, giving length and time the same dimension. In contrast, if we start from a three-base-unit system treating charge as a mechanical quantity, we cannot introduce an electrical fourth base unit, replacing some "1"s (which will not be visible) in an equation with "$4\pi\varepsilon_0$"s for example, without additional information. This is demonstrated below by the relative ease of changing from the SI to the Heaviside-Lorentz or Gaussian systems compared to the reverse procedures.

The extra base unit in the SI compared to CGS systems requires the existence of an extra dimensional constant, best identified as the 'electric constant' $\varepsilon_0$. This appears in the SI equation for the force between two charges in vacuum, $F = q_1 q_2 / 4\pi\varepsilon_0 r^2$, which can be taken as the definition of $\varepsilon_0$. The symbol $\mu_0$ – the 'magnetic constant' - is used as shorthand for $1/\varepsilon_0 c^2$, which appears in many equations describing magnetism.

The $4\pi$ appears in this equation because the divergence of the electric field is then given by div $\boldsymbol{E} = \rho/\varepsilon_0$, where $\rho$ is the charge density, whereas if we had used the equation $F = k_e q_1 q_2 / r^2$ (where $k_e$, the Coulomb constant, equals $1/4\pi\varepsilon_0$), the divergence would be div $\boldsymbol{E} = 4\pi k_e \rho$. The SI choice of where to put the $4\pi$ was called the "rationalised" one by Heaviside, who showed that this was the more natural place for it, because the $4\pi$ was then associated with spherical symmetry (e.g. [1] Section 2.2.3 pp. 413-414; [10] Appendix A.5-A.6; [16] Section 3.5.5).

The SI is called a rationalised unit system because its equations conventionally use $\varepsilon_0$ rather than $k_e$, but it can be "unrationalised" very easily, using the same equations but with $\varepsilon_0$ replaced by $1/4\pi k_e$, so that $\mu_0$ is replaced by $4\pi k_e/c^2$. There is no real difference between the rationalised and unrationalised SI equations, just a choice of how the constant is written down. Units, and the definitions of quantities, remain unchanged.

---

[2] Anyone who considers that electric charge is inextricably linked to force should remind themselves that it is central to electrochemistry as well as to electromagnetism. Michael Faraday measured charge accurately and objectively with his 'volta-electrometer' (later called a voltameter) back in 1833. The measurement does not involve forces on charges or currents. A justification of the minimum number of base units, starting from conservation principles, is given in [15].

It has not been necessary to mention any specific units in this section, because the "SI" equations are valid for any coherent set of units. In other words, any units can be used for the four base units provided that other units are derived from them according to the fundamental relationships between them, so that the charge unit is the current unit multiplied by the time unit, for example. The "SI" equations are better described as the "general" equations, which can readily accommodate being rationalised or unrationalised, and can use any coherent set of units, including an electrical unit that can be defined in terms of mechanical units by setting $\varepsilon_0$ or $\mu_0$ equal to a number. Of course, there are good reasons for using the metre, kilogram, second and ampere as the base units in practice, for global consistency, and to make use of the many derived units with special names.

Another conventional choice made within SI equations is to use the standard definition of magnetic flux density $B$, so that the magnetic Lorentz force is $F = qv \times B$. Alternatively, $B$ can be defined by $F = q(v/c) \times B$, but this is better described as a different quantity with a different symbol such as $B_R$ (for "relativistic B"), so that $B_R = cB$. $B_R$ has the same dimensions as the electric field $E$, which some people find appealing[3]. Of course, there is nothing to prevent SI equations being written in terms of $B_R$ instead of $B$, just by substituting $B_R/c$ for $B$ wherever it appears. $B_R$ would naturally go with a differently-defined magnetic vector potential $A_R$, where $A_R = cA$ so that $B_R = $ curl $A_R$, and a differently-defined magnetic flux, $\Phi_R$, where $\Phi_R = c\Phi$ so that $\Phi_R = B_R \times$ area.

Similarly, the magnetic quantities $H$ and $M$ conventionally used within SI equations can be substituted with $cH_R$ and $cM_R$. $H_R$ and $M_R$ have the same dimensions as $D$ and $P$, which also has some appeal. As $M$ is the magnetic dipole moment per unit volume, $M_R$ naturally goes with a differently-defined magnetic dipole moment $m_R$, where $m = cm_R$. Again, substitutions like these do not change the equations in any significant sense.

2.2 The Heaviside-Lorentz system

In the Heaviside-Lorentz system the constant $\varepsilon_0$ is removed from the SI equations by setting it equal to the number 1, so that $\mu_0$ becomes equal to $1/c^2$. Fundamentally, the conversion of SI equations to Heaviside-Lorentz equations can be done by replacing $\varepsilon_0$ and $\mu_0$ wherever they appear with 1 and $1/c^2$ respectively. The equation for the force between two charges becomes $F = q_1q_2/4\pi r^2$, so the system can be seen to be rationalised.

The less fundamental but more confusing feature is that Heaviside-Lorentz equations are always written in terms of $A_R$, $B_R$, $\Phi_R$, $H_R$, $m_R$ and $M_R$, where (as described in the previous section) $A_R = cA$, $B_R = cB$, $\Phi_R = c\Phi$, $H_R = H/c$, $m_R = m/c$ and $M_R = M/c$. The confusion arises because the subscripts are always omitted.

The removal of $\varepsilon_0$ has the effect of making the units for charge, mass, length and time inter-related. Any three of these four units can be chosen freely, and the fourth will be set by the requirement that the equation $F = q_1q_2/4\pi r^2$ holds true. Following the decision from 1873, the Heaviside-Lorentz system uses centimetres, grams and seconds for the three chosen base units. The Heaviside-Lorentz unit for charge has no special name but is equal to about 94.1 pC.

2.3 The Gaussian system

In the Gaussian system $\varepsilon_0$ is set equal to $1/4\pi$, the same as in the CGS-electrostatic system, so that $\mu_0$ is equal to $4\pi/c^2$. Fundamentally, the conversion of SI equations to Gaussian equations can be done by replacing $\varepsilon_0$ and $\mu_0$ wherever they appear with $1/4\pi$ and $4\pi/c^2$ respectively. The equation for the force between two charges becomes particularly simple, $F = q_1q_2/r^2$, demonstrating that the system is unrationalised.

---

[3] This is analogous to using the symbol $t_R$ to represent $ct$, where $t$ is a time, so that $t_R$ has the same dimensions as length. The natural place for $B$ to have the same dimensions as $E$ is within equations where $c$ has been set equal to 1.

The Gaussian system as currently used combines the two original CGS systems in a way that provides a single set of equations and a single set of units, favouring the electrical units from CGS-electrostatic and the magnetic units from CGS-electromagnetic. This is done by using CGS-electrostatic as the basis, while implicitly redefining some magnetic quantities along the same lines as in the Heaviside-Lorentz system.

Like Heaviside-Lorentz, Gaussian equations are always written in terms of the "relativistic" quantities $A_R$, $B_R$, $\Phi_R$, $m_R$ and $M_R$, described above. However, for historical reasons $H$ and $D$ in Gaussian equations have meanings that are different from both the SI and Heaviside-Lorentz versions. Gaussian equations are conventionally written in terms of $H_{RU}$ and $D_U$, where the subscript U denotes "unrationalised", and where $H_{RU} = 4\pi H_R$ and $D_U = 4\pi D$. Gaussian equations are also conventionally written in terms of unrationalised electrical and magnetic susceptibilities, $\chi_{eU}$ and $\chi_{mU}$, where $\chi_{eU} = \chi_e/4\pi$ and $\chi_{mU} = \chi_m/4\pi$. Again confusion arises because the subscripts are always omitted.

We have seen that the Gaussian equation for the force between two charges is $F = q_1q_2/r^2$, whereas the Heaviside-Lorentz version is $F = q_1q_2/4\pi r^2$. As the same (CGS) units for force and distance are used in both systems, we can see that because there is no constant in the equations that can absorb the effects of the rationalisation, the unit for charge, and indeed all electrical and magnetic units, must be different in the two systems. The Gaussian charge unit is $(4\pi)^{1/2}$ – about 3½ - times larger than the Heaviside-Lorentz one, and is called a statcoulomb, equal to about 0.334 nC.

## 3  Converting equations from SI to Heaviside-Lorentz or Gaussian

We can write down a simple, prescriptive method for converting any SI equation to the equation used in either the Heaviside-Lorentz or Gaussian systems, just by using the descriptions in Section 2. Tables 1 and 2 show which quantities need to be replaced, and include the subscripts that make it clear that some quantities have simply been substituted for others.

| SI symbol | $\varepsilon_0$ | $\mu_0$ | Magnetic vector potential $A$, magnetic flux $\Phi$ and magnetic flux density $B$ | Magnetic field strength $H$, magnetic dipole moment $m$ and magnetization $M$ |
|---|---|---|---|---|
| replacement in Heaviside-Lorentz | 1 | $1/c^2$ | $A_R/c$, $\Phi_R/c$, $B_R/c$ | $cH_R$, $cm_R$, $cM_R$ |

Table 1: Prescriptions for converting SI equations to the Heaviside-Lorentz system. The R subscripts highlight where the quantities are defined differently from the standard quantities used within the SI.

| SI symbol | $\varepsilon_0$ | $\mu_0$ | Magnetic vector potential $A$, magnetic flux $\Phi$ and magnetic flux density $B$ | Magnetic dipole moment $m$ and magnetization $M$ | Magnetic field strength $H$ | Electric displacement field $D$ | Susceptibility $\chi_e$ and $\chi_m$ |
|---|---|---|---|---|---|---|---|
| replacement in Gaussian | $1/4\pi$ | $4\pi/c^2$ | $A_R/c$, $\Phi_R/c$, $B_R/c$ | $cm_R$, $cM_R$ | $cH_{RU}/4\pi$ | $D_U/4\pi$ | $4\pi\chi_{eU}$, $4\pi\chi_{mU}$ |

Table 2: Prescriptions for converting SI equations to the Gaussian system. The R and U subscripts highlight where the quantities are defined differently from the standard quantities used within the SI.

Table 3 gives some examples of SI equations being converted to the other systems using the prescriptions given in Tables 1 and 2. Subscripts to distinguish different quantities are left visible, for clarity, though they are conventionally omitted.

| SI | Heaviside-Lorentz | Gaussian |
|---|---|---|
| $\boldsymbol{D} = \varepsilon_0 \boldsymbol{E} + \boldsymbol{P}$ | $\boldsymbol{D} = \boldsymbol{E} + \boldsymbol{P}$ | $\dfrac{\boldsymbol{D}_\text{U}}{4\pi} = \left(\dfrac{1}{4\pi}\right)\boldsymbol{E} + \boldsymbol{P}$ <br> $\boldsymbol{D}_\text{U} = \boldsymbol{E} + 4\pi \boldsymbol{P}$ |
| $\boldsymbol{H} = \dfrac{\boldsymbol{B}}{\mu_0} - \boldsymbol{M}$ | $c\boldsymbol{H}_\text{R} = \dfrac{\boldsymbol{B}_\text{R}}{c}(c^2) - c\boldsymbol{M}_\text{R}$ <br> $\boldsymbol{H}_\text{R} = \boldsymbol{B}_\text{R} - \boldsymbol{M}_\text{R}$ | $\dfrac{c\boldsymbol{H}_\text{RU}}{4\pi} = \dfrac{\boldsymbol{B}_\text{R}}{c}\left(\dfrac{c^2}{4\pi}\right) - c\boldsymbol{M}_\text{R}$ <br> $\boldsymbol{H}_\text{RU} = \boldsymbol{B}_\text{R} - 4\pi\boldsymbol{M}_\text{R}$ |
| $\text{div}\,\boldsymbol{E} = \rho/\varepsilon_0$ | $\text{div}\,\boldsymbol{E} = \rho$ | $\text{div}\,\boldsymbol{E} = 4\pi\rho$ |
| $\text{curl}\,\boldsymbol{B} = \mu_0\left(\boldsymbol{J} + \varepsilon_0 \dfrac{\partial \boldsymbol{E}}{\partial t}\right)$ | $\text{curl}\,\dfrac{\boldsymbol{B}_\text{R}}{c} = \left(\dfrac{1}{c^2}\right)\left(\boldsymbol{J} + \dfrac{\partial \boldsymbol{E}}{\partial t}\right)$ <br> $\text{curl}\,\boldsymbol{B}_\text{R} = \dfrac{1}{c}\left(\boldsymbol{J} + \dfrac{\partial \boldsymbol{E}}{\partial t}\right)$ | $\text{curl}\,\dfrac{\boldsymbol{B}_\text{R}}{c} = \left(\dfrac{4\pi}{c^2}\right)\left(\boldsymbol{J} + \left(\dfrac{1}{4\pi}\right)\dfrac{\partial \boldsymbol{E}}{\partial t}\right)$ <br> $\text{curl}\,\boldsymbol{B}_\text{R} = \dfrac{1}{c}\left(4\pi\boldsymbol{J} + \dfrac{\partial \boldsymbol{E}}{\partial t}\right)$ |

Table 3: Some examples of SI equations in electromagnetism being converted to the Gaussian and Heaviside-Lorentz versions using the method described here.

### 4      Converting equations from Heaviside-Lorentz or Gaussian to SI

Converting equations to SI from the Gaussian or Heaviside-Lorentz versions is more complicated than the other way around. It is simple to revert to the standard (SI) definitions of the differently-defined quantities, using the relations given in sections 2.2 and 2.3. However, after $\varepsilon_0$ has been replaced with a number, reinstating it in the right places requires extra information, such as from dimensional analysis (e.g. [16]). Specifically, we can insert $\varepsilon_0$ (when converting from Heaviside-Lorentz), or $4\pi\varepsilon_0$ (when converting from Gaussian), wherever this is necessary to balance the dimensions within a four-base-unit system. This asymmetry in converting between the unit systems should not be surprising once it is appreciated that the process of conversion from SI to Gaussian or Heaviside-Lorentz equations loses some information.

Full dimensional analysis of equations using a system of four dimensions such as mass **M**, length **L**, time **T** and charge **Q** would be quite tedious. Fortunately, because we know we are only looking to reinstate any missing '$\varepsilon_0$'s, which have the dimensions **M**$^{-1}$ **L**$^{-3}$ **T**$^2$ **Q**$^2$, we need only consider one of these dimensions, as the other three can be assumed to take care of themselves. While it would seem natural to consider the charge dimension **Q**, it is more convenient to consider the mass dimension **M**.

Using the set of four dimensions **M**, **L**, **T** and **Q**, all commonly-encountered physical quantities that have a mass dimension have one of either **M**$^1$ or **M**$^{-1}$. The operators grad, div and curl, and multiplication or division by $c$ to create the "relativistic" versions of quantities described above, change the length and time dimensions of the relevant quantities, but do not change their mass dimension. They can therefore be ignored when considering the mass dimension. The charge dimension is not changed by these operations either, but physical quantities can have charge dimensions of **Q**$^{-2}$ (e.g. resistance), **Q**$^{-1}$ (e.g. voltage), **Q** (e.g. current) or **Q**$^2$ (e.g. capacitance), and this makes the process more complicated. Specifically, in this case factors of $\varepsilon_0^{1/2}$ or $(4\pi\varepsilon_0)^{1/2}$ need to be inserted into equations during conversion, as well as factors of $\varepsilon_0$ or $4\pi\varepsilon_0$, even though no square-root terms appear in the final converted equations.

Full physical dimensions for many quantities can be seen in, for example, [16] Appendix 1. The most relevant physical quantities are presented within the mass dimension categories in Table 4.

| Mass dimension | Physical quantity |
|---|---|
| Dimensionless quantities | relative permittivity $\varepsilon_r$ <br> relative permeability $\mu_r$ <br> susceptibility $\chi_e$ and $\chi_m$ |
| Other quantities with no mass dimension | charge $q$, charge density $\rho$, and electric displacement field **D** (charge and its distribution) <br> current $I$, current density **J**, and magnetic field strength **H** (current and its distribution) <br> electric dipole moment **p** (charge x length), and polarization **P** (e-dipole density) <br> magnetic dipole moment **m** (current x area) and magnetization **M** (m-dipole density) <br> length, time, frequency, velocity, acceleration, angle, angular velocity, area, and volume. |
| **M**$^1$ | force **F**, torque $\tau$, energy $E$ and power $P$ <br> (also mass, action, momentum, moment of inertia and angular momentum) <br> voltage or potential difference $V$ (energy/charge) <br> electric flux $\phi_E$ and electric field **E** ('force/charge') <br> magnetic flux $\Phi$, magnetic flux density **B** and magnetic vector potential **A** ('force/current') <br> resistance $R$ and resistivity $\rho$ ('voltage/current') <br> inductance $L$ (magnetic flux/current) <br> permeability e.g. $\mu_0$ |
| **M**$^{-1}$ | conductance $G$ and conductivity $\sigma$ ('current/voltage') <br> capacitance $C$ (charge/voltage) <br> permittivity e.g. $\varepsilon_0$ |

Table 4: the mass dimension of most physical quantities likely to be encountered in electromagnetic equations. Both the names and conventional symbols for each quantity are given, for clarity. Note that the symbol $\rho$ is conventionally used for both charge density and resistivity.

Any missing '$\varepsilon_0$'s (when converting from Heaviside-Lorentz) or '$4\pi\varepsilon_0$'s (from Gaussian) can be seen from Table 4 to have a mass dimension of **M**$^{-1}$. We can therefore determine whether and where we need to insert $\varepsilon_0$ or $4\pi\varepsilon_0$ by seeing whether a factor of **M**$^{-1}$ is needed to balance the mass dimension within an equation. Equivalently, if a factor of **M** is needed, we insert $1/\varepsilon_0$ or $1/(4\pi\varepsilon_0)$. Any $\varepsilon_0$ can then be replaced by $1/\mu_0 c^2$ if desired. Unlike other published methods (e.g. [13] and [14]), factors of $\varepsilon_0^{1/2}$ or $(4\pi\varepsilon_0)^{1/2}$ do not feature.

For example, within the Gaussian version of the Coulomb force equation, $F = q_1 q_2/r^2$, only one quantity has a mass dimension, the force $F$, with mass dimension **M**. We can therefore make the equation dimensionally balanced by inserting a factor of $4\pi\varepsilon_0$ next to $F$. Similarly, for the Heaviside-Lorentz version, $F = q_1 q_2/4\pi r^2$, we can make the equation dimensionally balanced by inserting a factor of $\varepsilon_0$ next to $F$. In both cases[4] the equation converts to SI as $F = q_1 q_2/4\pi\varepsilon_0 r^2$, as shown in Tables 6 and 8.

While the insertions can often be done by inspection, it is possible to write prescriptive methods for introducing the factors, shown in Tables 5 and 7, which combine the dimensional considerations with the different definitions used within the different systems previously described. Tables 6 and 8 give examples.

## 5 Converting equations between the Heaviside-Lorentz and Gaussian systems

Prescriptions for converting equations between the Heaviside-Lorentz and Gaussian systems can be found by directly combining the earlier results, and are given in Table 9, for completeness.

---

[4] It might feel odd to insert a constant having an electrical dimension ($\varepsilon_0$) to accompany a mechanical quantity like force. This does not change the dimensions of force, however; it is purely to balance dimensions within the equation as a whole.

## 6    Comments

Enthusiasts for the Gaussian system like the fact that their equations never include $\varepsilon_0$ or $\mu_0$, and that their **B** has the same dimensions as **E**. They tend not to mention that, being unrationalised, "4π"s appear in strange places, as shown in the examples above. It is a great pity that the Gaussian system was invented and widely used before the appearance of the Heaviside-Lorentz system, which has the same benefits as the Gaussian system without the rationalisation problem. The Gaussian system has continued to overshadow the Heaviside-Lorentz one purely because of familiarity and inertia, and because some of its units have special names.

If the Gaussian system could be quietly dropped, there would be no need to surreptitiously introduce the quantities $H_{RU}$, $D_U$, $\chi_{eU}$ and $\chi_{mU}$, and students could progress directly from the SI to the Heaviside-Lorentz system much more easily. Better still, students could be taught that the alternative set of definitions for **B**, **H** and **M** can be used perfectly well within the SI, and the Heaviside-Lorentz system would simply be the SI with $\varepsilon_0$ set equal to 1.

| Symbol in Heaviside-Lorentz equations | magnetic vector potential **A**, magnetic flux $\Phi$ and magnetic flux density **B** | magnetic field strength **H**, magnetic dipole moment **m** and magnetization **M** | force **F**, torque $\tau$, energy E, power P, voltage V, electric flux $\phi_E$, electric field **E**, resistance R, resistivity $\rho$ and inductance L | conductance G, conductivity $\sigma$ and capacitance C |
|---|---|---|---|---|
| **Conversion to SI** | multiply by $\varepsilon_0 c$ | divide by $c$ | multiply by $\varepsilon_0$ | divide by $\varepsilon_0$ |

Table 5: Prescriptions for converting equations in the Heaviside-Lorentz systems to SI equations. Subscripts are omitted as they will not generally appear in Heaviside-Lorentz equations as found.

| Heaviside-Lorentz | SI | Heaviside-Lorentz | SI |
|---|---|---|---|
| $\boldsymbol{D} = \boldsymbol{E} + \boldsymbol{P}$ | $\boldsymbol{D} = \varepsilon_0 \boldsymbol{E} + \boldsymbol{P}$ | $P = IV$ | $\varepsilon_0 P = I\varepsilon_0 V$ <br> $P = IV$ |
| $\boldsymbol{H} = \boldsymbol{B} - \boldsymbol{M}$ | $\boldsymbol{H}/c = \varepsilon_0 c \boldsymbol{B} - \boldsymbol{M}/c$ <br> $\boldsymbol{H} = \varepsilon_0 c^2 \boldsymbol{B} - \boldsymbol{M}$ <br> $\boldsymbol{H} = \boldsymbol{B}/\mu_0 - \boldsymbol{M}$ | $C = \varepsilon_r A/d$ | $C/\varepsilon_0 = \varepsilon_r A/d$ <br> $C = \varepsilon_0 \varepsilon_r A/d$ |
| $\operatorname{div} \boldsymbol{E} = \rho$ | $\operatorname{div} \varepsilon_0 \boldsymbol{E} = \rho$ <br> $\operatorname{div} \boldsymbol{E} = \rho/\varepsilon_0$ | $\boldsymbol{F} = q\boldsymbol{E} + \dfrac{q\boldsymbol{v}}{c} \times \boldsymbol{B}$ | $\varepsilon_0 \boldsymbol{F} = q\varepsilon_0 \boldsymbol{E} + \dfrac{q\boldsymbol{v}}{c} \times \varepsilon_0 c\boldsymbol{B}$ <br> $\boldsymbol{F} = q\boldsymbol{E} + q\boldsymbol{v} \times \boldsymbol{B}$ |
| $\operatorname{curl} \boldsymbol{B} = \dfrac{1}{c}\left(\boldsymbol{J} + \dfrac{\partial \boldsymbol{E}}{\partial t}\right)$ | $\varepsilon_0 c \operatorname{curl} \boldsymbol{B} = \dfrac{1}{c}\left(\boldsymbol{J} + \varepsilon_0 \dfrac{\partial \boldsymbol{E}}{\partial t}\right)$ <br> $\operatorname{curl} \boldsymbol{B} = \mu_0 \left(\boldsymbol{J} + \varepsilon_0 \dfrac{\partial \boldsymbol{E}}{\partial t}\right)$ | $F = \dfrac{q_1 q_2}{4\pi r^2}$ | $\varepsilon_0 F = \dfrac{q_1 q_2}{4\pi r^2}$ <br> $F = \dfrac{q_1 q_2}{4\pi \varepsilon_0 r^2}$ |

Table 6: Examples of equations being converted from the Heaviside-Lorentz system to SI. Subscripts are omitted.

| Symbol in Gaussian equations | magnetic vector potential **A**, magnetic flux $\Phi$ and magnetic flux density **B** | magnetic field strength **H** | magnetic dipole moment **m** and magnetization **M** | electric displacement field **D** | susceptibility $\chi_e$ and $\chi_m$ |
|---|---|---|---|---|---|
| **Conversion to SI** | multiply by $4\pi\varepsilon_0 c$ | multiply by $4\pi/c$ | divide by $c$ | multiply by $4\pi$ | divide by $4\pi$ |

| Symbol in Gaussian equations (continued) | force **F**, torque $\tau$, energy $E$, power $P$, voltage $V$, electric flux $\phi_E$, electric field **E**, resistance $R$, resistivity $\rho$ and inductance $L$ | conductance $G$, conductivity $\sigma$ and capacitance $C$ |
|---|---|---|
| **Conversion to SI** | multiply by $4\pi\varepsilon_0$ | divide by $4\pi\varepsilon_0$ |

Table 7: Prescriptions for converting equations in the Gaussian system to SI equations. Subscripts are omitted as they will not generally appear in Gaussian equations as found.

| Gaussian | SI | Gaussian | SI |
|---|---|---|---|
| $\mathbf{D} = \mathbf{E} + 4\pi\mathbf{P}$ | $4\pi\mathbf{D} = 4\pi\varepsilon_0\mathbf{E} + 4\pi\mathbf{P}$ <br> $\mathbf{D} = \varepsilon_0\mathbf{E} + \mathbf{P}$ | $P = IV$ | $4\pi\varepsilon_0 P = I4\pi\varepsilon_0 V$ <br> $P = IV$ |
| $\mathbf{H} = \mathbf{B} - 4\pi\mathbf{M}$ | $4\pi\mathbf{H}/c = 4\pi\varepsilon_0 c\mathbf{B} - 4\pi\mathbf{M}/c$ <br> $\mathbf{H} = \varepsilon_0 c^2 \mathbf{B} - \mathbf{M}$ <br> $\mathbf{H} = \mathbf{B}/\mu_0 - \mathbf{M}$ | $C = \varepsilon_r A/4\pi d$ | $C/4\pi\varepsilon_0 = \varepsilon_r A/4\pi d$ <br> $C = \varepsilon_0 \varepsilon_r A/d$ |
| $\operatorname{div} \mathbf{E} = 4\pi\rho$ | $4\pi\varepsilon_0 \operatorname{div} \mathbf{E} = 4\pi\rho$ <br> $\operatorname{div} \mathbf{E} = \rho/\varepsilon_0$ | $\mathbf{F} = q\mathbf{E} + \dfrac{q\mathbf{v}}{c} \times \mathbf{B}$ | $4\pi\varepsilon_0 \mathbf{F} = q4\pi\varepsilon_0 \mathbf{E} + \dfrac{q\mathbf{v}}{c} \times 4\pi\varepsilon_0 c\mathbf{B}$ <br> $\mathbf{F} = q\mathbf{E} + q\mathbf{v} \times \mathbf{B}$ |
| $\operatorname{curl} \mathbf{B} = \dfrac{1}{c}\left(4\pi\mathbf{J} + \dfrac{\partial \mathbf{E}}{\partial t}\right)$ | $4\pi\varepsilon_0 c \operatorname{curl} \mathbf{B} = \dfrac{1}{c}\left(4\pi\mathbf{J} + 4\pi\varepsilon_0 \dfrac{\partial \mathbf{E}}{\partial t}\right)$ <br> $\operatorname{curl} \mathbf{B} = \mu_0 \left(\mathbf{J} + \varepsilon_0 \dfrac{\partial \mathbf{E}}{\partial t}\right)$ | $F = \dfrac{q_1 q_2}{r^2}$ | $4\pi\varepsilon_0 F = \dfrac{q_1 q_2}{r^2}$ <br> $F = \dfrac{q_1 q_2}{4\pi\varepsilon_0 r^2}$ |

Table 8: Examples of equations in the Gaussian system being converted to the SI. Subscripts are omitted.

| Quantity and symbol | magnetic vector potential **A**, magnetic flux $\Phi$, magnetic flux density **B**, magnetic field strength **H**, electric displacement field **D**, electric field **E**, voltage $V$, electric flux $\phi_E$, resistance $R$, resistivity $\rho$, inductance $L$, force **F**, torque $\tau$, energy $E$ and power $P$ | conductance $G$, conductivity $\sigma$, capacitance $C$, susceptibility $\chi_e$ and $\chi_m$ |
|---|---|---|
| **Conversion to Heaviside-Lorentz** | multiply by $4\pi$ | divide by $4\pi$ |
| **Conversion to Gaussian** | divide by $4\pi$ | multiply by $4\pi$ |

Table 9: Prescriptions for converting equations between the Gaussian and Heaviside-Lorentz systems